\documentclass{webofc}

\usepackage[varg]{txfonts}   
\usepackage{xcolor}
\usepackage{hyperref}
\usepackage{url}
\usepackage{comment}
\usepackage{braket}
\hypersetup{colorlinks=true,citecolor=blue,urlcolor=blue,linkcolor=blue}

\begin{document}
%
\title{\textit{Ab initio}-guided optical potentials for semi-magic nuclei from Gorkov self-consistent Green's functions}
\author{\firstname{Thibault} \lastname{Demartini}\inst{1,2}\fnsep\thanks{\email{thibault.demartini@cea.fr}} \and
        \firstname{Guillaume} \lastname{Blanchon}\inst{1,2} \and
        \firstname{Vittorio} \lastname{Som\`a}\inst{3} \and
        \firstname{Thomas} \lastname{Duguet}\inst{3}
}

\institute{CEA, DAM, DIF, 91297 Arpajon, France
\and
           Universit\'e Paris-Saclay, CEA, Laboratoire Mati\`ere en Conditions Extr\^emes, 91680 Bruy\`eres-Le-Ch\^atel, France
\and
           Universit\'e Paris-Saclay, CEA, IRFU, 91191 Gif-sur-Yvette, France
          }

\abstract{We construct \textit{ab initio}-guided nonlocal optical potentials for semi-magic nuclei from the self-consistent Gorkov-ADC(2) one-body Green's function. This extends \textit{ab initio} self-consistent Green's function approaches to open-shell targets and to proton projectiles. The starting point is the NNLO$_{\rm sat}$ chiral interaction, whose three-nucleon component is included as an effective two-body interaction. The self-energy is computed in a harmonic-oscillator basis, transformed to coordinate space and used in the scattering code SIDES. Its imaginary part relies on a phenomenological energy-dependent width, hence the label \textit{ab initio}-guided. Differential and integral (total and reaction) cross sections are compared with data for n,p$+^{40}$Ca and p$+^{42}$Ca. Forward angles are well reproduced. Backward angles reveal a lack of absorption, which calls for larger model spaces and many-body truncations beyond Gorkov-ADC(2).}
\maketitle
%
\section{Introduction}
\label{sec:intro}

The optical potential is the standard tool to reduce the nucleon-nucleus scattering problem to an effective one-body problem. It enters an integro-differential Schr\"odinger equation whose solution gives access to elastic scattering observables. Deriving it from first principles serves several purposes. (i) It would provide a microscopic framework to interpret scattering measurements. (ii) It would allow predictions for exotic nuclei far from stability, where no data exist. (iii) It would guide global phenomenological parametrizations. (iv) It would enable the propagation of uncertainties from the underlying chiral interaction to reaction observables.

Traditionally, optical potentials are built empirically: a functional form is assumed a priori and its parameters are fitted to experimental data~\cite{Varner1991,Koning2003,Perey1962}. Such potentials are very successful within their calibration domain. However, extrapolation beyond their calibration range is uncontrolled, and their predictive power for exotic nuclei is limited. This motivates the shift toward microscopic optical potentials.
A modern route is the dispersive optical model (DOM)~\cite{Mahaux1991}, extended to nonlocal potentials by Dickhoff, Charity and collaborators~\cite{Dickhoff2019}. The DOM treats the optical potential as a complex, nonlocal and dispersive self-energy. It is constrained simultaneously by scattering and structure data, including along the Ca isotopic chain~\cite{Charity2007}. The DOM is, however, fitted to data, whereas the present approach aims at computing the self-energy from the nuclear Hamiltonian.

Microscopic strategies fall into two broad categories. The first folds an effective interaction over the target density. This interaction can be either a $g$ matrix~\cite{Arellano1989} or a $t$ matrix~\cite{Vorabbi2016}. These approaches are best suited to intermediate and high incident energies, namely $\gtrsim50~\rm MeV$ ($\gtrsim100~\rm MeV$) if in-medium effects are included (absent) in the effective interaction.
At lower energies, the structure of the target is not negligible but is not captured by these approximations. The second strategy, which we follow here, directly computes the target self-energy in the finite nucleus via an \textit{ab initio} many-body method based on chiral nucleon-nucleon (NN) and three-nucleon (3N) interactions. Within self-consistent Green's function (SCGF) theory, the irreducible self-energy is identified with the optical potential~\cite{BellSquires1959}; its relation to Feshbach's optical potential~\cite{Feshbach1958} is discussed in Refs.~\cite{Capuzzi1996,Waldecker2011}. This route offers a systematic path to improvement and, in principle, predictive power for exotic nuclei. Several many-body schemes have been pursued, based on the algebraic diagrammatic construction (ADC) of the self-energy~\cite{Idini2019}, on coupled-cluster theory~\cite{Rotureau2017} or, more recently, on the generator coordinate method~\cite{Bostrom2026}.

A key previous application within the ADC framework is the work of Idini, Barbieri and Navr\'atil~\cite{Idini2019}, in which neutron elastic scattering off $^{16}$O and $^{40}$Ca was computed from a Dyson-ADC(3) self-energy. The comparison with low-energy data was satisfactory. The Dyson formalism is, however, restricted to closed-shell (doubly-magic) targets, and the application was limited to neutron projectiles.

In the present work, we lift both restrictions. The Gorkov extension of SCGF theory~\cite{Soma2011,Soma2014} breaks $U(1)$ symmetry associated with particle-number conservation in order to capture pairing correlations. It thus gives access to the much larger set of semi-magic nuclei. We use it at the Gorkov-ADC(2) level for open-shell targets, and we treat both neutron and proton projectiles. The self-energy is computed from the NNLO$_{\rm sat}$ interaction~\cite{Ekstrom2015} employing a harmonic-oscillator (HO) basis. It is then transformed to coordinate space and used as a nonlocal optical potential in the scattering code SIDES~\cite{Blanchon2020}. Because its imaginary part presently relies on a phenomenological width (Sec.~\ref{sec:gorkov}), we refer to the resulting potentials as \textit{ab initio}-guided rather than \textit{ab initio}.


\section{Gorkov-ADC(2) optical potential}
\label{sec:gorkov}

To compute the optical potential, we start from the intrinsic Hamiltonian
\begin{equation}
    H_{\rm int} = {T} - {T}_{\rm c.m.} + {V} + {W} \, , 
\end{equation}
where ${T}_{\rm c.m.}$ is the center-of-mass kinetic energy, $V$ denotes the NN potential including strong and Coulomb interactions, and $W$ represents the 3N force.
The latter is included as an effective two-body interaction, obtained by averaging the 3N force over the one-body density~\cite{Carbone2013}.
In the present application, $V$ and $W$ are taken from the NNLO$_{\rm sat}$ Hamiltonian of Ref.~\cite{Ekstrom2015}.

Gorkov SCGF theory starts from the grand potential $\Omega = H_{\rm int} - \mu_p{Z} - \mu_n {N}$, where ${Z}$ (${N}$) is the proton (neutron) number operator. The chemical potentials $\mu_p$ and $\mu_n$ are adjusted such that the proton and neutron numbers are constrained to their target values on average, $\langle Z\rangle = \bra{\Psi_0}{Z}\ket{\Psi_0}$ and $\langle N\rangle = \bra{\Psi_0}{N}\ket{\Psi_0}$. In this symmetry-broken treatment, the reference state $\ket{\Psi_0}$ does not carry a definite particle number.

Since $U(1)$ symmetry is broken, anomalous contractions of two annihilation (or two creation) operators no longer vanish. It is then natural to work in Nambu space, with the spinor $(a_a, \bar{a}_a^\dagger)^{T}$, where $\bar{a}$ annihilates the time-reversed state. The dynamics is embodied in the corresponding 2$\times$2 Gorkov propagator
\begin{equation}
    G(E) = \begin{pmatrix}
        G^{11}(E) & G^{12}(E)\\
        G^{21}(E) & G^{22}(E)
    \end{pmatrix}\,,
\end{equation}
whose static limit is the Hartree-Fock-Bogoliubov (HFB) problem. Its blocks read, in the Lehmann representation,

\begin{align}
\label{eq:gorkov_propagators}
G^{11}_{ab}(E) &= \sum_{k}
  \left[ \frac{\mathcal{U}^{k}_{a}\,\mathcal{U}^{k*}_{b}}{E-E_{k}+i\eta}
       + \frac{\bar{\mathcal{V}}^{k*}_{a}\,\bar{\mathcal{V}}^{k}_{b}}{E+E_{k}-i\eta} \right] ,
\end{align}
whereas $G^{12}$, $G^{21}$ and $G^{22}$ follow by exchanging $U\leftrightarrow V$ and applying time reversal \cite{Soma2011,Soma2014}. Here, Latin subscripts $a$ and $b$ label states of the one-body HO basis, and a bar denotes the corresponding time-reversed state. The index $k$ runs over the eigenstates $\ket{\Psi_k}$ of $\Omega$, with eigenvalues $\Omega_k$. The poles are $E_{k} \equiv \Omega_{k} - \Omega_{0}$. They are thus measured with respect to the chemical potentials, rather than being excitation energies at fixed particle number. One-nucleon addition and removal energies are recovered as $E^{\pm}_{k} = \mu_{q_{k}} \pm E_{k}$, with $q_{k}$ the isospin projection carried by the pole. The infinitesimal $\eta \to 0^{+}$ fixes the time-ordering boundary conditions. The residues are built from the amplitudes $\mathcal{U}^{k}$ and $\mathcal{V}^{k}$ to reach $\ket{\Psi_{k}}$ by adding a nucleon to, or removing one from, $\ket{\Psi_{0}}$ in a given single-particle state. The normal propagator $G^{11}$ contains both the addition ($\mathcal{U}$) and removal ($\bar{\mathcal{V}}$) parts of the one-nucleon strength distribution; $G^{22}$ is its time-reversed counterpart. The anomalous propagators $G^{12}$ and $G^{21}$ have no counterpart in Dyson's formalism. They encode the pairing correlations arising from the breaking of particle-number symmetry, and vanish in the closed-shell limit. The four propagators are solutions of the Gorkov equation, the $2\times2$ matrix generalization of Dyson's equation. The normal [$\Sigma^{11}$, $\Sigma^{22}$] and anomalous [$\Sigma^{12}$, $\Sigma^{21}$] irreducible self-energies act as energy-dependent potentials.

Following Refs.~\cite{Capuzzi1996,Waldecker2011,Idini2019}, the optical potential is identified with an ADC(2) self-energy in Dyson form, evaluated with the poles and residues of the Gorkov normal propagator $G^{11}$. 
It is computed in the laboratory frame. For a given partial wave $(l,j)$, the self-energy is transformed to coordinate space through the HO radial functions,
\begin{equation} 
    \label{eq:op} 
    \Sigma^{lj}(r,r',E) = \sum_{n_a n_b} \phi^{\rm HO}_{n_a l}(r)\, \Sigma_{n_a n_b }^{lj}(E)\, \phi^{\rm HO}_{n_b l}(r')\,.
\end{equation} 
It is then fed to the scattering code SIDES~\cite{Blanchon2020} to compute the elastic scattering observables. The HO basis is truncated at $e_{\max} \equiv \max(2n+l)$, where $n$ and $l$ are the radial and orbital quantum numbers. We use an oscillator frequency $\hbar\omega=20$~MeV, chosen such that the binding energy is minimized with respect to $\hbar\omega$ at the largest $e_{\max}$~\cite{Soma2014}. The calculations employ $e_{\max}=7$ and $11$ for $^{40}$Ca, and $e_{\max}=11$ for $^{42}$Ca.

In a finite HO basis, the continuum is discretized. The self-energy of Eq.~(\ref{eq:op}) then consists of a set of isolated poles, and its imaginary part reduces to a sum of delta functions in the limit $\eta\to 0^+$. To obtain a smooth self-energy suitable for scattering calculations, we follow the prescription of Refs.~\cite{Idini2019,Bostrom2026}. The infinitesimal $\eta$ in the pole denominators of the dynamical part of the self-energy is replaced by an energy-dependent width applied to $\Sigma$ after its self-consistent determination,
\begin{equation}
\eta \;\rightarrow\; \Gamma(E) = \frac{1}{\pi}\,\frac{a\,(E-E_F)^2}{(E-E_F)^2+b^2}\,.
\label{eq:lorentzian}
\end{equation}
Here $E_F$ is the Fermi energy, $b=22.36$~MeV, and $a$ typically ranges from 12 to 24~MeV~\cite{Bostrom2026}. As in Ref.~\cite{Bostrom2026}, we checked that the observables shown below are weakly sensitive to $a$ within this range, and we adopt $a=24$~MeV. Note that the finite width smooths the real part of the dynamical self-energy as well as its imaginary part.

Replacing $\eta$ by a finite width amounts to an energy average of the self-energy, which is the standard way of defining the optical potential~\cite{Capuzzi1996}. The magnitude and energy dependence of $\Gamma(E)$ are, however, phenomenological. They mimic the spreading width due to configurations beyond ADC(2) and to missing fragmentation. They also mimic the escape width due to the coupling to the continuum, which is absent from a bound HO basis. The $(E-E_F)^2$ behavior accounts for the reduction of phase space near the Fermi energy. 
\section{Results}
\label{sec:results}

In this section, the cross sections obtained with the present approach for neutron and proton scattering off $^{40}$Ca and $^{42}$Ca are compared to experimental data. We start with neutron scattering off $^{40}$Ca. We then show that the proton channel is affected by the truncation of the Coulomb interaction, and we propose a minimal correction. Finally, first results for p$+^{42}$Ca scattering are presented.

\subsection{Differential, total and reaction cross sections for n$+^{40}$Ca}
\label{sec:diffn}

\begin{figure}[t]
\centering
\includegraphics[width=0.9\linewidth]{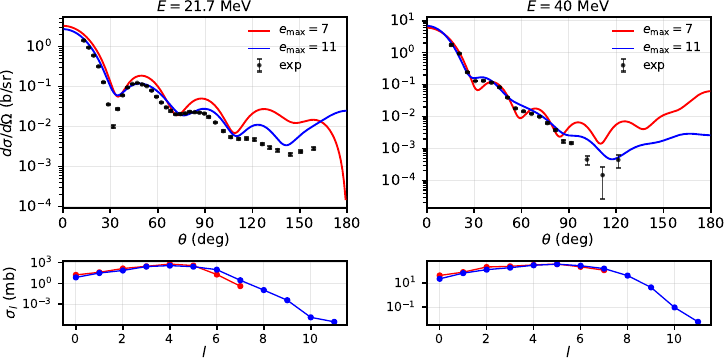}
\caption{Differential cross sections and partial cross sections $\sigma_l$ [Eq.~(\ref{eq:sigmal})] for n$+^{40}$Ca at incident neutron energies of 21.7 and 40~MeV (laboratory frame). The Gorkov-ADC(2) potential (NNLO$_{\rm sat}$) is used with $e_{\max}=7$ (red) and $e_{\max}=11$ (blue). Data from Refs.~\cite{Alarcon1987} (21.7~MeV) and~\cite{DeVito1981} (40~MeV).}
\label{fig:neutron40Ca}
\end{figure}

Figure~\ref{fig:neutron40Ca} displays the differential cross sections for neutron elastic scattering off $^{40}$Ca at 21.7 and 40~MeV. They are computed with $e_{\max}=7$ and $11$ and compared with the experimental data of Refs.~\cite{Alarcon1987,DeVito1981}. The figure also shows the partial cross sections $\sigma_l$, used as a diagnostic of convergence in partial waves. They are defined as
\begin{equation}
\sigma_l = \frac{2\pi}{k^2}\sum_{j}(2j+1)|T_{lj}|^2\,,
\label{eq:sigmal}
\end{equation}
where $k$ is the wave number and $T_{lj}$ the nuclear $T$-matrix element, so that $\sum_l \sigma_l = \sigma_{\rm el}$ is the elastic cross section.

Since the self-energy is built from HO states with $2n+l\le e_{\max}$, it vanishes identically for $l>e_{\max}$. These partial waves are therefore not scattered by the nuclear potential. The model space must thus contain enough partial waves at a given energy. At 40~MeV, the partial cross section of the last partial wave, $l=e_{\max}$, is still of order $10^{1}$~mb for $e_{\max}=7$. It drops to $10^{-2}$~mb for $e_{\max}=11$. The smaller model space is therefore not converged in partial waves at this energy, while the larger one nearly is. Higher energies would require larger values of $e_{\max}$. 

With $e_{\max}=11$, the ADC(2) potential reproduces the measured angular distributions well at forward angles at both energies, although the first minimum at 21.7~MeV is too shallow. The $e_{\max}=7$ calculation is clearly insufficient; at 21.7~MeV its drop at the most backward angles is a further sign of its lack of convergence. At larger angles, the $e_{\max}=11$ calculation overestimates the data, by up to an order of magnitude. Together with the too shallow first minimum, this is the typical signature of an absorption that is insufficient or incorrectly distributed in radius. It is consistent with correlations missing at the ADC(2) level, such as the resummation of ring (RPA-like) and ladder diagrams in the 2p1h space. 
Below about 15~MeV (not shown), the agreement deteriorates further. 

\begin{figure}[t]
\centering
\includegraphics[width=0.9\linewidth]{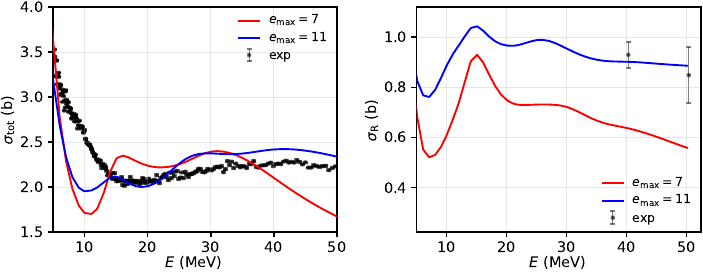}
\caption{Total (left) and reaction (right) cross sections for n$+^{40}$Ca as a function of the incident neutron energy (laboratory frame), computed with the Gorkov-ADC(2) potential. They are compared with the total~\cite{Finlay1993} and nonelastic~\cite{Zanelli1981} cross section data (see text for the distinction between reaction and nonelastic cross sections).}
\label{fig:neutron40CaTot}
\end{figure}

The total and reaction cross sections are shown in Fig.~\ref{fig:neutron40CaTot} for incident energies between 5 and 50~MeV. Three regions can be distinguished. Below about 15~MeV, both truncations fail to reproduce the data. As for the angular distributions, this is consistent with missing correlations and with the sparseness of the discretized spectrum at low energy; the oscillations of $\sigma_{\rm R}$ in this region reflect the latter. Between 15 and 35~MeV, the $e_{\max}=11$ calculation follows the measured total cross section. Its reaction cross section passes through the two available data points, whereas $e_{\max}=7$ underestimates them. Above 35~MeV, the $e_{\max}=7$ total cross section drops sharply, reaching 1.6~b at 50~MeV against 2.25~b in the data. This is again due to the partial-wave truncation: at 50~MeV, the partial waves carrying most of the cross section are absent from the $e_{\max}=7$ model space. The $e_{\max}=11$ result remains stable over the whole energy range.
The optical-model reaction cross section $\sigma_{\rm R}$ measures the flux removed from the shape-elastic channel. It therefore includes compound-elastic scattering, which is not part of the measured nonelastic cross section: $\sigma_{\rm R} = \sigma_{\rm non} + \sigma_{\rm CE}$. At the energies of the two data points, around 40 and 50~MeV, compound-elastic scattering is negligible for $^{40}$Ca, so that the comparison is meaningful.

\subsection{Differential cross sections for p$+^{40}$Ca}
\label{sec:diffp}

\begin{figure}[t]
\centering
\includegraphics[width=0.9\linewidth]{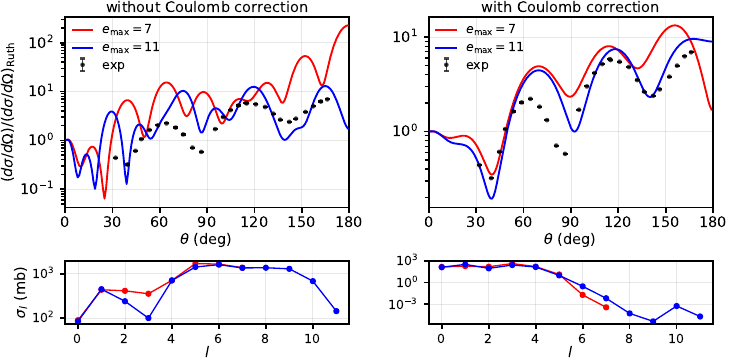}
\caption{Ratio of the differential cross section to the Rutherford cross section, and partial cross sections $\sigma_l$, for p$+^{40}$Ca at 15.65~MeV (laboratory frame). Left: without the Coulomb correction. Right: with the Coulomb correction (see text). Data from Ref.~\cite{Lombardi1972}.}
\label{fig:proton40Ca}
\end{figure}

The proton channel requires a separate discussion. In the many-body calculation, the Coulomb interaction between protons is treated on the same footing as the nuclear interaction. Its matrix elements are computed in the truncated HO basis and added to the NN$+$3N ones~\cite{Soma2011,Soma2014}. The direct Coulomb contribution to the self-energy then inherits the Gaussian fall-off of the HO functions. It lacks the long-range $1/r$ tail, even at $e_{\max}=11$. Without correction, the direct Coulomb term is thus truncated. This shows up as partial cross sections that do not vanish at large $l$ (left panel of Fig.~\ref{fig:proton40Ca}).

A first minimal correction is presented. The direct Coulomb term, represented in the truncated HO basis, is subtracted from the static self-energy. The long-range Coulomb interaction is then accounted for by the local Coulomb potential of SIDES, a uniformly charged sphere, which has the correct $1/r$ asymptotics. This avoids representing the $1/r$ tail with HO functions. The Coulomb exchange term and the Coulomb contributions to the dynamical self-energy are kept in the HO representation.

The corrected results are shown in the right panel of Fig.~\ref{fig:proton40Ca}. The partial cross sections now decrease with $l$ as expected. The $e_{\max}=7$ and $e_{\max}=11$ calculations coincide up to $l\approx 5$. The ratio to the Rutherford cross section is well reproduced up to about $60^\circ$, including the position of the first diffraction minimum around $40^\circ$. The second minimum, observed around $80^\circ$, is not reproduced. The calculation overestimates the maximum around $60^\circ$ and yields a shallower minimum, shifted to larger angles. At backward angles, the oscillation pattern is in phase with the data, but its amplitude is overestimated. As in the neutron case, this points to insufficient absorption. At this relatively low energy, compound-elastic contributions to the backward-angle data may also play a role.

\subsection{First results for p$+^{42}$Ca}
\label{sec:Ca42}

So far, the Gorkov features of the method have not been tested. Since $^{40}$Ca is doubly magic, the anomalous components of the Green's function and of the self-energy vanish. We now turn to $^{42}$Ca: the proton shell $Z=20$ is closed, while the two valence neutrons in the $0f_{7/2}$ shell are paired. Quadrupole correlations beyond the spherical, paired description are not included. Since the proton shell is closed, neutron pairing enters the proton optical potential only indirectly, through intermediate configurations involving neutron quasiparticles. 

\begin{figure}[t]
\centering
\includegraphics[width=0.9\linewidth]{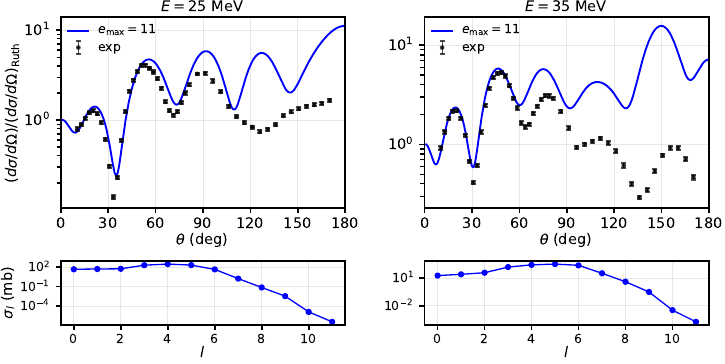}
\caption{Ratio of the differential cross section to the Rutherford cross section, and partial cross sections $\sigma_l$, for p$+^{42}$Ca at 25 and 35~MeV (laboratory frame), computed with $e_{\max}=11$. Data from Ref.~\cite{McCamis1986}.}
\label{fig:proton42Ca}
\end{figure}

The differential cross sections for proton scattering off $^{42}$Ca at 25 and 35~MeV are displayed in Fig.~\ref{fig:proton42Ca}. As shown by the partial cross sections, the calculations are converged in partial waves. The same conclusion as for $^{40}$Ca can be drawn: the data at forward angles are well reproduced, but the calculation fails to reproduce the backward-angle cross section.


\section{Conclusions and outlook}
\label{sec:conclusion}

We have presented a first application of Gorkov self-consistent Green's function self-energies to nucleon scattering off a semi-magic target. The resulting \textit{ab initio}-guided optical potentials are based on the Gorkov-ADC(2) approximation with the NNLO$_{\rm sat}$ chiral interaction. They have been applied to both neutron and proton projectiles. For n$+^{40}$Ca, differential cross sections are well reproduced at forward angles. With $e_{\max}=11$, total and reaction cross sections follow the data between about 15 and 50~MeV. For protons, replacing the HO-represented direct Coulomb term restores consistency with the Coulomb asymptotics. Forward angles are then well reproduced for p$+^{40}$Ca and p$+^{42}$Ca. In all cases, backward angles are overestimated, which points to insufficient absorption.

The basis truncation $e_{\max}$ limits the accessible incident energies, while the ADC(2) truncation is likely the main source of missing absorption; particle-number restoration is not included either. We therefore plan to go to Gorkov-ADC(2)-X, which adds first-order interactions among the 2p1h and 2h1p configurations~\cite{Schirmer1983} and recovers part of the ring and ladder resummations, and then to Gorkov-ADC(3)~\cite{Barbieri2022}. The dependence on the HO basis ($\hbar\omega$, $e_{\max}$) and on the chiral interaction will also be quantified.

Neutron elastic scattering off $^{42}$Ca and heavier open-shell isotopes, where neutron pairing enters the optical potential directly, would offer a more direct test, but no such data are currently available. In this context, one should keep in mind that current realistic chiral interactions, such as the one employed here, have been shown to produce too weak pairing correlations~\cite{Soma2021,Scalesi2026}, so that the impact of pairing on low-energy scattering is presumably underestimated in the present calculations.

\section*{Acknowledgements}

This project was provided with HPC and storage resources by GENCI at TGCC, France, thanks to Grant No. A0190513012 on the supercomputer Joliot-Curie’s ROME partition.




\begin{thebibliography}{99}

\bibitem{Varner1991}
R.L. Varner \textit{et al.}, Phys. Rep. \textbf{201}, 57 (1991). \href{https://doi.org/10.1016/0370-1573(91)90039-O}{doi:10.1016/0370-1573(91)90039-O}

\bibitem{Koning2003}
A.J. Koning, J.P. Delaroche, Nucl. Phys. A \textbf{713}, 231 (2003). \href{https://doi.org/10.1016/S0375-9474(02)01321-0}{doi:10.1016/S0375-9474(02)01321-0}

\bibitem{Perey1962}
F. Perey, B. Buck, Nucl. Phys. \textbf{32}, 353 (1962). \href{https://doi.org/10.1016/0029-5582(62)90345-0}{doi:10.1016/0029-5582(62)90345-0}

\bibitem{Mahaux1991}
C. Mahaux, R. Sartor, Adv. Nucl. Phys. \textbf{20}, 1 (1991). \href{https://doi.org/10.1007/978-1-4613-9910-0_1}{doi:10.1007/978-1-4613-9910-0\_1}


\bibitem{Dickhoff2019}
W.H. Dickhoff, R.J. Charity, Prog. Part. Nucl. Phys. \textbf{105}, 252 (2019). \href{https://doi.org/10.1016/j.ppnp.2018.11.002}{doi:10.1016/j.ppnp.2018.11.002}

\bibitem{Charity2007}
R.J. Charity \textit{et al.}, Phys. Rev. C \textbf{76}, 044314 (2007). \href{https://doi.org/10.1103/PhysRevC.76.044314}{doi:10.1103/PhysRevC.76.044314}

\bibitem{Arellano1989}
H.F. Arellano, F.A. Brieva, W.G. Love, Phys. Rev. Lett. \textbf{63}, 605 (1989). \href{https://doi.org/10.1103/PhysRevLett.63.605}{doi:10.1103/PhysRevLett.63.605}

\bibitem{Vorabbi2016}
M. Vorabbi, P. Finelli, C. Giusti, Phys. Rev. C \textbf{93}, 034619 (2016). \href{https://doi.org/10.1103/PhysRevC.93.034619}{doi:10.1103/PhysRevC.93.034619}

\bibitem{BellSquires1959}
J.S. Bell, E.J. Squires, Phys. Rev. Lett. \textbf{3}, 96 (1959). \href{https://doi.org/10.1103/PhysRevLett.3.96}{doi:10.1103/PhysRevLett.3.96}

\bibitem{Feshbach1958}
H. Feshbach, Ann. Phys. (N.Y.) \textbf{5}, 357 (1958). \href{https://doi.org/10.1016/0003-4916(58)90007-1}{doi:10.1016/0003-4916(58)90007-1}

\bibitem{Capuzzi1996}
F. Capuzzi, C. Mahaux, Ann. Phys. (N.Y.) \textbf{245}, 147 (1996). \href{https://doi.org/10.1006/aphy.1996.0006}{doi:10.1006/aphy.1996.0006}

\bibitem{Waldecker2011}
S.J. Waldecker, C. Barbieri, W.H. Dickhoff, Phys. Rev. C \textbf{84}, 034616 (2011). \href{https://doi.org/10.1103/PhysRevC.84.034616}{doi:10.1103/PhysRevC.84.034616}

\bibitem{Idini2019}
A. Idini, C. Barbieri, P. Navr\'atil, Phys. Rev. Lett. \textbf{123}, 092501 (2019). \href{https://doi.org/10.1103/PhysRevLett.123.092501}{doi:10.1103/PhysRevLett.123.092501}

\bibitem{Rotureau2017}
J. Rotureau \textit{et al.}, Phys. Rev. C \textbf{95}, 024315 (2017). \href{https://doi.org/10.1103/PhysRevC.95.024315}{doi:10.1103/PhysRevC.95.024315}

\bibitem{Bostrom2026}
J. Bostr\"om, B.G. Carlsson, A. Idini, arXiv:2604.01033 [nucl-th] (2026).

\bibitem{Soma2011}
V. Som\`a, T. Duguet, C. Barbieri, Phys. Rev. C \textbf{84}, 064317 (2011). \href{https://doi.org/10.1103/PhysRevC.84.064317}{doi:10.1103/PhysRevC.84.064317}

\bibitem{Soma2014}
V. Som\`a, C. Barbieri, T. Duguet, Phys. Rev. C \textbf{89}, 024323 (2014). \href{https://doi.org/10.1103/PhysRevC.89.024323}{doi:10.1103/PhysRevC.89.024323}

\bibitem{Ekstrom2015}
A. Ekstr\"om \textit{et al.}, Phys. Rev. C \textbf{91}, 051301(R) (2015). \href{https://doi.org/10.1103/PhysRevC.91.051301}{doi:10.1103/PhysRevC.91.051301}

\bibitem{Blanchon2020}
G. Blanchon \textit{et al.}, Comput. Phys. Commun. \textbf{254}, 107340 (2020). \href{https://doi.org/10.1016/j.cpc.2020.107340}{doi:10.1016/j.cpc.2020.107340}

\bibitem{Carbone2013}
A. Carbone \textit{et al.}, Phys. Rev. C \textbf{88}, 054326 (2013). \href{https://doi.org/10.1103/PhysRevC.88.054326}{doi:10.1103/PhysRevC.88.054326}


\bibitem{Alarcon1987}
R. Alarcon, J. Rapaport, R.W. Finlay, Nucl. Phys. A \textbf{462}, 413 (1987). \href{https://doi.org/10.1016/0375-9474(87)90397-6}{doi:10.1016/0375-9474(87)90397-6}

\bibitem{DeVito1981}
R.P. DeVito \textit{et al.}, Phys. Rev. Lett. \textbf{47}, 628 (1981). \href{https://doi.org/10.1103/PhysRevLett.47.628}{doi:10.1103/PhysRevLett.47.628}

\bibitem{Finlay1993}
R.W. Finlay \textit{et al.}, Phys. Rev. C \textbf{47}, 237 (1993). \href{https://doi.org/10.1103/PhysRevC.47.237}{doi:10.1103/PhysRevC.47.237}

\bibitem{Zanelli1981}
C.I. Zanelli \textit{et al.}, Phys. Rev. C \textbf{23}, 1015 (1981). \href{https://doi.org/10.1103/PhysRevC.23.1015}{doi:10.1103/PhysRevC.23.1015}

\bibitem{Lombardi1972} J.C. Lombardi, R.N. Boyd, R. Arking, A.B. Robbins,
Nucl. Phys. A \textbf{188}, 103 (1972). doi:10.1016/0375-9474(72)90186-8

\bibitem{McCamis1986}
R.H. McCamis \textit{et al.}, Phys. Rev. C \textbf{33}, 1624 (1986). \href{https://doi.org/10.1103/PhysRevC.33.1624}{doi:10.1103/PhysRevC.33.1624}

\bibitem{Schirmer1983}
J. Schirmer, L.S. Cederbaum, O. Walter, Phys. Rev. A \textbf{28}, 1237 (1983). \href{https://doi.org/10.1103/PhysRevA.28.1237}{doi:10.1103/PhysRevA.28.1237}


\bibitem{Barbieri2022}
C. Barbieri, T. Duguet, V. Som\`a, Phys. Rev. C \textbf{105}, 044330 (2022). \href{https://doi.org/10.1103/PhysRevC.105.044330}{doi:10.1103/PhysRevC.105.044330}

\bibitem{Soma2021} 
V. Som\`a \textit{et al.}, Eur. Phys. J. A \textbf{57}, 135 (2021). \href{https://doi.org/10.1140/epja/s10050-021-00437-4}{doi:10.1140/epja/s10050-021-00437-4}

\bibitem{Scalesi2026} 
A. Scalesi \textit{et al.}, Phys. Rev. C \textbf{113}, L051303 (2026). \href{doi:10.1103/mmy4-3wrp}{doi:10.1103/mmy4-3wrp}

\end{thebibliography}
\end{document}